\documentclass[11p]{article}
\usepackage{graphicx}
\usepackage{amsmath}
\usepackage{amssymb}
\usepackage[numbers,compress]{natbib}
\begin{document}

\title{Integral representation of the scalar propagators on the de Sitter expanding universe}

\author{Ion I.  Cot\u aescu\thanks{E-mail:~~~i.cotaescu@e-uvt.ro} ,
and Ion Cot\u aescu Jr. \thanks{E-mail:~~~ion.cotaescu@e-uvt.ro}\\
{\small \it West University of Timi\c soara,}\\
{\small \it V.  P\^ arvan Ave.  4, RO-300223 Timi\c soara, Romania}}

\maketitle

\begin{abstract}
A new type of integral representation is proposed for the propagators of the massive Klein-Gordon field minimally coupled to the gravity of the de Sitter expanding universe. As a simple application the amplitudes of the Compton effect in the second order of perturbations are derived.

 Pacs:
04.62.+v
\end{abstract}


\newpage

\section{Introduction}

The classical or quantum scalar fields are the principal pieces used in various models on curved spacetime.  Of a special interest in cosmology is the de Sitter expanding universe carrying scalar fields variously coupled to gravity whose quantum modes can be analytically solved \cite{Nach,CT,BODU,T,Csc,Csc1}. Nevertheless, despite of this opportunity, we have not yet a complete scalar quantum field theory (QFT) on de Sitter backgrounds based on perturbations and renormalization procedures able to describe all the processes involving scalar bosons. This is because of the technical difficulties in calculating Feynman diagrams affecting the fields of any spin on the de Sitter expanding universe.

The source of these difficulties is the fact that the causal propagators, expressed explicitly in terms of Heaviside step functions depending on time, lead to the fragmentation of the time integrals of the chronological products of free fields giving the transition amplitudes in different orders of perturbations.  Under such circumstances, these integrals cannot be evaluated forcing one to restrict so far only to the first order amplitudes of the de Sitter QFT  which do not involve propagators \cite{Lot1,Lot2,Lot3,R1,R2,R3,AuSp1,AuSp2,CQED,Cr1,Cr2}.  Note that the processes in the first order of perturbations which are forbidden in special relativity by the energy-momentum conservation are allowed on the de Sitter spacetimes where the momentum and energy cannot be measured simultaneously \cite{CGRG}. However, the calculations in the first order of perturbations are only the first step to a complete QFT involving propagators for which we must get over the above mentioned difficulties.

In the traditional  QFT on the Minkowskian spacetime this problem is solved by the Fourier representations of the causal propagators which encapsulate the effect of the Heaviside step functions \cite{BDR}. Unfortunately, in the de Sitter case such Fourier representations do not hold as we shall explain in what follows. Therefore, we must look for another type of integral representation able to take over the effects of the Heaviside functions. Recently we succeeded to find a new integral representation  of the propagators of the Dirac  field on the de Sitter \cite{Cint1} or any spatially flat FLRW \cite{Cint2} spacetimes   which is different from the usual Fourier integrals allowed in special relativity.  Here we would like to continue this study applying the same method to the massive and charged Klein-Gordon  fields, minimally coupled to the gravity of the de Sitter expanding universe, writing down for the first time the new integral representation of their propagators. 

Moreover, we show that this new integral representation plays the same role as the familiar Fourier one in special relativity, helping us to calculate the Feynman diagrams of the de Sitter scalar quantum electrodynamics (SQED) in a similar manner as in the flat case. In order to convince that we present as a premiere  the amplitudes of the Compton effect on the de Sitter expanding universe,  written in a closed form thanks to our integral representation.

We start in the second section presenting briefly the massive scalar field whose mode functions are written in the conformal local chart with Cartesian coordinates. The next section is devoted to our principal result reported here, demonstrating that the integral representation we propose gives just the Feynman propagator after applying the method of contour integrals \cite{BDR}. In the third section we derive for the first time the amplitudes of the Compton effect on the de Sitter expanding portion. Finally some concluding remarks are presented.   

\section{Massive scalar field}

Let us start with the de Sitter expanding universe defined as the expanding portion of the $(3+1)$-dimensional de Sitter manifold, equipped with the spatially flat FLRW chart whose coordinates, $x^{\mu}$ ($\alpha,...\mu,\nu,...=0,1,2,3$), are the proper time $x^0=t$ and the Cartesian coordinates $x^i$ ($i,j,k,...=1,2,3$) for which we use the vector notation, ${\bf x}=(x^1,x^2,x^3)$. For technical reasons we work here mainly in the conformal  chart having the conformal time 
\begin{equation}\label{tc}
t_c=-\frac{1}{\omega}e^{-\omega t}<0\,,
\end{equation}
and the same space coordinates.  In these charts the line element reads  \cite{BD}
\begin{equation}\label{mrw}
ds^2=g_{\mu\nu}(x)dx^{\mu}dx^{\nu}=dt^2-a(t)^2(d{\bf x}\cdot d{\bf x})= a(t_c)^2(dt_c^2-d{\bf x}\cdot d{\bf x})\,,
\end{equation}
where 
\begin{equation}\label{atc}
a(t)=e^{\omega t} \to a(t_c)=-\frac{1}{\omega t_c}\,,
\end{equation}
depend on the Hubble constant of the de Sitter spacetime denoted here by $\omega$. 

In the conformal chart, the Klein-Gordon equation of a charged scalar particle of mass $m$  takes the form 
\begin{equation}\label{KG1}
\left( \partial_{t_c}^2-\Delta -\frac{2}{t_c}\,
\partial_{t_c}+\frac{m^2}{\omega^2 t_c^2}\right)\phi(x_c)=0\,,
\end{equation}
allowing the well-known solutions  that can be expanded in terms of plane waves of
positive and negative frequencies as \cite{BD}
\begin{equation}\label{field1}
\phi(x)=\phi^{(+)}(x)+\phi^{(-)}(x)=\int d^3p \left[f_{\bf p}(x)a({\bf
p})+f_{\bf p}^*(x)b^*({\bf p})\right]
\end{equation}
where the fundamental solutions have the general form 
\begin{equation}
f_{\bf p}(x)=f_{\bf p}(t_c,{\bf x})=\frac{1}{\sqrt{\pi\omega}}\frac{1}{[2\pi a(t_c)]^{\frac{3}{2}}} {\cal F}_{\nu}(t_c) \, e^{i {\bf p}\cdot {\bf x}}\,.
\end{equation}
The time modulation function ${\cal F}_{\mu}$ may be any arbitrary linear combination of Bessel functions. In what follows it is convenient to consider modified Bessel functions $K_{\nu}$ instead of the usual Hankel ones  such that we can write
\begin{equation}
{\cal F}_{\nu}(t_c)=\alpha K_{i\nu}(ipt_c)+\beta K_{i\nu}(-ipt_c)\,,
\end{equation}
where $\nu=\sqrt{\frac{m^2}{\omega^2}-\frac{9}{4}}$ (in the minimal coupling) 
while $\alpha$ and $\beta$ are arbitrary complex valued arbitrary constants. On the other hand, the fundamental solutions must  satisfy the orthonormalization relations
\begin{eqnarray}
\langle  f_{\bf p},f_{{\bf p}'}\rangle=-\langle  f_{\bf p}^*,f_{{\bf
p}'}^*\rangle&=&\delta^3({\bf p}-{\bf p}')\,,\\
\langle f_{\bf p},f_{{\bf p}'}^*\rangle&=&0\,,
\end{eqnarray}
with respect to the relativistic scalar product \cite{BD}
\begin{equation}\label{SP2}
\langle \phi,\phi'\rangle=i\int \frac{d^3x}{(\omega t_c)^2}\, \phi^*(x)
\stackrel{\leftrightarrow}{\partial_{t_c}} \phi'(x)\,,
\end{equation}
that hold only if we set $|\alpha|^2-|\beta|^2=1$,  as it results from Eqs. (\ref{SP2}) and (\ref{KuKu}).

The quantization can be done in a canonical manner by fixing the constants $\alpha$ and $\beta$ for determining the vacuum \cite{P1} and replacing then the wave
functions of the field (\ref{field1}) with field operators, $a({\bf p}) \to {\frak a}({\bf p})$ and  
$b({\bf p}) \to {\frak b}({\bf p})$, such that $b^{*}\to {\frak  b}^{\dagger}$  \cite{BDR}. Here we chose the Bunch-Davies vacuum \cite{BuD}, with $\alpha=1$ and $\beta=0$, which is a particular case of adiabatic vacuum \cite{GM,Zel,ZelS,GMM,allen,bousso}. Furthermore, we assume  that the particle (${\frak a}$, ${\frak a}^{\dagger}$) and antiparticle (${\frak b}$, ${\frak b}^{\dagger}$) operators fulfill the standard commutation relations in the momentum representation, among which the non-vanishing ones are
\begin{equation}\label{com1}
[a({\bf p}), a^{\dagger}({\bf p}^{\,\prime})]=[b({\bf p}), b^{\dagger}({\bf
p}^{\,\prime})] = \delta^3 ({\bf p}-{\bf p}^{\,\prime})\,.
\end{equation}
In the configurations representation the partial commutator functions of positive or negative frequencies, 
\begin{equation}
iD^{(\pm)}(x,x')=[\phi^{(\pm)}(x),\phi^{(\pm)\,\dagger}(x')]
\end{equation}
give the total one, $D=D^{(+)}+D^{(-)}$. These functions are solutions of the Klein-Gordon
equation in both the sets of variables and obey $[D^{(\pm)}(x,x')]^*=D^{(\mp)}(x,x')$  such that $D$ is a real valued function. These functions can be written as mode integrals as, 
\begin{eqnarray}
&&iD^{(+)}(x,x')=iD^{(+)}(t_c,t_c',{\bf x}-{\bf x}')=\int d^3 p \, f_{\bf p}(x)f_{\bf p}(x')^* \nonumber\\
&&~~~~~~=\frac{1}{\pi\omega}\frac{1}{[4\pi^2 a(t_c)a(t_c')]^{\frac{3}{2}}} \int d^3 p\,e^{i{\bf p}\cdot({\bf x}-{\bf x}')}K_{i\nu}(ipt_c)K_{i\nu}(-ipt_c')\,,\label{Dp}\\
&&iD^{(-)}(x,x')=iD^{(-)}(t_c,t_c',{\bf x}-{\bf x}')=-\int d^3 p \, f_{\bf p}(x)^*f_{\bf p}(x') \nonumber\\
&&~~~~~~=-\frac{1}{\pi\omega}\frac{1}{[4\pi^2 a(t_c)a(t_c')]^{\frac{3}{2}}} \int d^3 p\,e^{i{\bf p}\cdot({\bf x}-{\bf x}')}K_{i\nu}(-ipt_c)K_{i\nu}(ipt_c')\,,\label{Dm}
\end{eqnarray}
taking similar forms after changing ${\bf p}\to -{\bf p}$ in the last integral.  Note that these integrals can be solved in terms of hypergeometric functions obtaining well-known closed formulas \cite{BD}.  

\section{Propagators}

The commutator functions  allow us to construct the propagators, i. e. the Green functions corresponding to extreme initial conditions, without solving the Green equation. As  in the scalar theory on Minkowski spacetime, we may use the Heaviside step functions for defining   the retarded, $D_R$, and advanced, $D_A$, propagators,  
\begin{eqnarray}
D_R(t_c,t_c',{\bf x}-{\bf x}')&=& \theta(t_c-t_c')D(t_c,t_c',{\bf x}-{\bf x}')\,,\\
D_A(t_c,t_c',{\bf x}-{\bf x}')&=& -\,\theta(t_c'-t_c)D(t_c,t_c',{\bf x}-{\bf x}')\,,
\end{eqnarray}
while the causal Feynman propagator has the well-known form \cite{BDR},
\begin{eqnarray}\label{DF}
&&iD_F(t_c,t_c',{\bf x}-{\bf x}')= \langle 0|T[\phi(x)\phi^{\dagger}(x')]\,|0\rangle
\nonumber\\
&&~~~~= \theta (t_c-t_c') D^{(+)}(t_c,t_c',{\bf x}-{\bf x}')-\theta(t_c'-t_c)D^{(-)}(t_c,t_c',{\bf
x}-{\bf x}')\,.
\end{eqnarray}
Our main goal here is to find  suitable integral representations of these propagators which should encapsulate the effect of the Heaviside step functions. 

{ \begin{figure}
  \centering
    \includegraphics[scale=0.40]{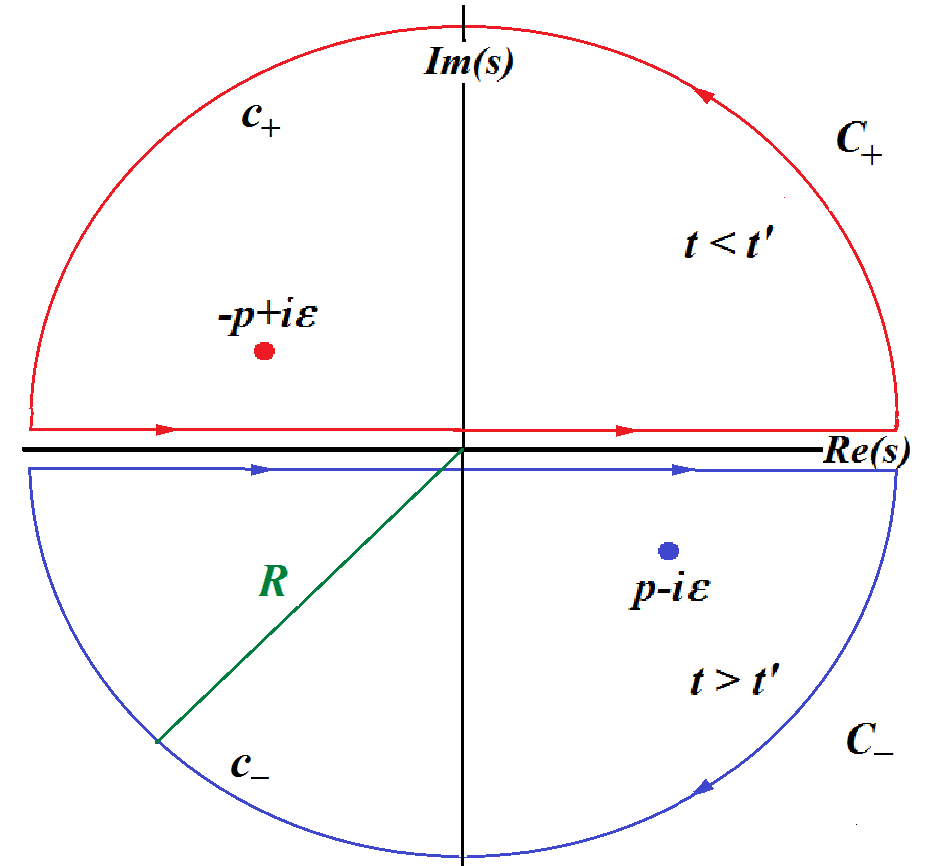}
    \caption{The contours of integration in the complex $s$-plane, $C_{\pm}$, are the limits of the pictured ones for $ R\to \infty$.}
  \end{figure}}

The explicit forms of the partial commutator functions (\ref{Dp}) and (\ref{Dm}) suggest us to postulate the following integral representation of the Feynman propagator
\begin{eqnarray}
D_F(x,x')&\equiv& D_F(t_c,t_c',{\bf x}-{\bf x}')=\frac{1}{\pi^2\omega}\frac{1}{[4\pi^2 a(t_c)a(t_c')]^{\frac{3}{2}}} \nonumber\\
&\times& \int d^3p\, e^{i{\bf p}\cdot({\bf x}-{\bf x}')}\int_{-\infty}^{\infty} ds\, |s| \,\frac{K_{i\nu}(ist_c)K_{i\nu}(-ist_c')}{s^2-p^2-i\epsilon}\,. \label{DFI}
\end{eqnarray}
It remains to prove that this integral representation gives just the Feynman propagator (\ref{DF}) according to the well-known method of contour integrals \cite{BDR}. Focusing on the last integral of Eq. (\ref{DFI}) denoted as  
\begin{equation}
{\cal I}(t_c,t_c')=\int_{-\infty}^{\infty}ds\,M(s,t_c,t_c')\,,
\end{equation}
we observe that for  large values of $|s|$ the modified Bessel functions can be approximated as in Eqs. (\ref{Km0}) obtaining the asymptotic  behavior  
\begin{equation}
M(s,t_c,t_c')\sim \frac{e^{-is(t_c-t_c')}}{s}\,. 
\end{equation}
Now we can  estimate the integrals on the semicircular parts, $c_{\pm}$, of the contours pictured in Fig. 1 taking $s\sim R e^{i\varphi}$ and using Eq. (3.338-6) of Ref. \cite{GR} which gives
\begin{equation}
\int_{c_{\pm}}ds\,M(s,t_c,t_c')\sim  I_0[\pm R(t_c-t_c')]\sim \frac{1}{\sqrt{R}}\,e^{\pm R(t_c-t_c')}\,,
\end{equation}
since the modified Bessel function $I_0$ behaves as in the first of Eqs. (\ref{Km0}). In the limit of $R\to \infty$ the contribution of the semicircle $c_+$ vanishes for $t_c'>t_c$ while those of the semicircle $c_-$ vanishes for $t_c>t_c'$. Therefore, the integration along the real $s$-axis  is equivalent with the following contour integrals
\begin{equation}
{\cal I}(t_c,t_c')=\left\{
\begin{array}{lll}
\int_{\small C_+}ds\,M(s,t_c,t_c')={\cal I}_+(t_c,t_c')&{\rm for}& t_c<t_c'\\
\int_{\small C_-}ds\,M(s,t_c,t_c')={\cal I}_-(t_c,t_c')&{\rm for}&t_c>t_c'
\end{array}\right. \,,\nonumber
\end{equation} 
where the contours $C_{\pm}$ are the limits for $R\to \infty$ of those of Fig. 1. Then we  may apply  the Cauchy's theorem \cite{Complex}, 
\begin{equation}
{\cal I}_{\pm}(t_c,t_c')=\pm 2\pi i \left.{\rm Res}\left[M(s,t_c,t_c')\right]\right|_{s=\mp p\pm i\epsilon}\,,
\end{equation}
taking into account that in the simple poles at $s=\pm p\mp i\epsilon$ we have the residues
\begin{equation}
\left.{\rm Res}\left[M(s,t,t')\right]\right|_{s=\pm  p\mp i\epsilon}=\pm\frac{1}{2}\,K_{\nu}(\pm ip t_c){K}_{\nu}(\mp i pt'_c)\,.
\end{equation}
Consequently,  the integral ${\cal I}_-(t_c,t_c')$ gives the first term of the Feynman propagator (\ref{DF}) with $D^{(+)}$ expanded as in Eq. (\ref{Dp}) while the integral ${\cal I}_+(t_c,t_c')$ yields its second term with $D^{(-)}$ in the form (\ref{Dm}), proving that the integral rep. (\ref{DFI}) is correct.  

The other propagators, $D_A$ and $D_R$, can be represented in a similar manner by changing the positions of the poles as in the flat case \cite{BDR},
\begin{eqnarray}
{D}_{\substack{R\\
A}}(x,x')&=&{D}_{\substack{R\\
A}}(t_c,t_c^{\prime},{\bf x}-{\bf x}')
=\frac{1}{\pi^2\omega} \frac{1}{[4\pi^2 a(t_c) a(t_c')]^{\frac{3}{2}}}\nonumber\\
&\times&\int {d^3p}\,{e^{i{\bf p}\cdot({\bf x}-{\bf x}')}}\int_{-\infty}^{\infty}ds\, |s|\,\frac{K_{i\nu}(ist_c)K_{i\nu}(-ist_c')}{(s\pm i\epsilon)^2-p^2}\,,\label{SRA}
\end{eqnarray}  
but in our integral representation instead of the Fourier one. 

Finally we note that the above integral representations can be rewritten at any time in the FLRW chart, $\{t,{\bf x})\}$, substituting $t_c\to t$ and $a(t_c)\to a(t)$ according to Eqs. (\ref{tc}) and (\ref{atc}).

\section{Compton effect in SQED}

We succeeded thus to derive the specific integral representations of the scalar propagators on the de Sitter  expanding universe that can be used  for calculating the Feynman  diagrams of the physical effects involving the Klein-Gordon field. Here we would like to give a simple example outlining how our approach works in the SQED on the de Sitter expanding universe,  deriving the amplitudes of the Compton effect in the second order of perturbations. 

We consider that our massive charged scalar field $\phi$  is coupled minimally to the electromagnetic field $A_{\mu}$ through the interaction Lagrangian
\begin{equation}
{\cal L}_{int}=-i e \sqrt{g(x)}\,g^{\mu\nu}(x)A_{\mu}(x)\left[\phi^{\dagger}(x)\stackrel{\leftrightarrow}{\partial_{\nu}}\phi(x)\right]\,,
\end{equation}
where $e$ is the electrical charge. We know that in the chart $\{t_c,{\bf x}\}$ the electromagnetic potential can be expanded in terms of similar plane waves as in the Minkowski spacetime since the Maxwell equations are conformally invariant if we work exclusivelly in the Coulomb gauge where $A_0=0$ \cite{CMax,CQED}. Therefore, we may write the expansion
\begin{equation}\label{Max}
{A_i}(x)=\int d^3k
\sum_{\lambda}\left[{\mu}_{{\bf k},\lambda;\,i}(x) \alpha({\vec
k},\lambda)+{\mu}_{{\bf k},\lambda;\,i}(x)^* \alpha^{\dagger}({\bf k},\lambda)\right]\,,
\end{equation}
in terms of the mode functions,
\begin{equation}\label{fk}
{\mu}_{{\bf k},\lambda;\,i}(t_c,{\bf x}\,)=
\frac{1}{(2\pi)^{3/2}}\frac{1}{\sqrt{2k}}\,e^{-ikt_c+i{\bf
k}\cdot {\bf x}}\,{\varepsilon}_{i} ({\bf k},\lambda)\,,
\end{equation}
depending on the components of the polarization vectors ${\varepsilon}_{i}({\bf k},\lambda)$ of momentum ${\bf k}$ ($k=|{\bf k}|$) and helicity $\lambda=\pm 1$ \cite{CQED}.   Note that the polarization vector is orthogonal to momentum, $k^i\varepsilon_{i}({\bf k},\lambda)=0$.

With these preparations we can write the first Compton amplitude \cite{BDR} with a self-explanatory notation as
\begin{eqnarray}
&&{\cal A}_{\lambda_1,\lambda_2}({\bf p}_1, {\bf k}_1, {\bf p}_2, {\bf k}_2)\equiv\langle out\,{\bf p}_2, ({\bf k}_2,\lambda_2)|in \,{\bf p}_1, ({\bf k}_1,\lambda_1\rangle \nonumber\\ 
&&=  -i \,\frac{e^2}{2} \int d^4x\,d^4x' \sqrt{g(x)g(x')}\,g^{ij}(x)g^{kl}(x') \mu_{{\bf k}_2,\lambda_2, j}(x)^*\mu_{{\bf k}_1,\lambda_1, l}(x')\nonumber\\
&&~~~\times\left[f_{{\bf p}_2}^*(x)  \stackrel{\leftrightarrow}{\partial_{i}} D_F(x,x') \stackrel{\leftrightarrow}{\partial_{k}'}f_{{\bf p}_1}(x') \right]\,,
\end{eqnarray}
where $D_F$ is given by our integral representation  (\ref{DFI}). Ve perform first the space integrals generating Dirac $\delta$-functions which have to assure the momentum conservation after integrating over the internal momentum ${\bf p}$ of $D_F$. Thus, after a little calculation we may write
\begin{eqnarray}
{\cal A}_{\lambda_1,\lambda_2}({\bf p}_1, {\bf k}_1, {\bf p}_2, {\bf k}_2)&=&-i\frac{e^2}{2}\,\delta^3({\bf p}_1+{\bf k}_1-{\bf p}_2-{\bf k}_2)p_1^i\varepsilon_i({\bf k}_1,\lambda_1)p_2^j\varepsilon_j({\bf k}_2,\lambda_2)^*\nonumber\\
&\times & \int_{-\infty}^{\infty} ds |s| \frac{{\cal V}(p_2,k_2,s)^*{\cal V}(p_1,k_1,s)}{s^2-|{\bf p}_1 +{\bf k}_1|^2-i\epsilon}\,,\label{A1}
\end{eqnarray} 
where we introduced the vertex functions defined up to a phase factor as
\begin{equation}\label{vert}
{\cal V}(p,k,s)=\frac{1}{2\omega^2 \pi^3 \sqrt{k}}\int_0^{\infty}d\tau\,\tau\, K_{i\nu}\left(-i\frac{p}{\omega}\tau\right)K_{i\nu}\left(i\frac{s}{\omega}\tau\right)e^{i\frac{k}{\omega}\tau}
\end{equation} 
after changing the variable of integration $t_c\to \tau=-\omega t_c$. We obtained thus a closed form of the first Compton amplitude bearing in mind that the second amplitude can be obtained directly by changing ${\bf k}_1 \leftrightarrow {\bf k}_2$ in Eq. (\ref{A1})  \cite{BDR}. 

We must stress that this result could not be obtained without our integral representation since the Heaviside step functions of the original form (\ref{DF}) mix up the time integrals. However, the  Compton amplitudes obtained here are complicated since, in general, the quantum effects in de Sitter spacetimes are described by formulas involving integrals of the form (\ref{vert}). For example, in the de Sitter QED in Coulomb gauge \cite{CQED} the amplitudes in the first order of perturbations are given by integrals of this form whose analyze required an extended analytical and numerical study \cite{CQED}.  A similar study can be performed in the case of the Compton effect calculated here but this exceeds the purposes of this paper where the principal objective was to define our new integral representation.   
 
\section{Concluding remarks}

The above example shows that the integral representation of the scalar propagators proposed here is crucial for calculating the Feynman diagrams in any order of the SQED in the presence of the gravity of the de Sitter background. Thus one could find  new observable effects involving interacting fields, allowed by the local gravity, whose indirect influence could be better measured than its direct interaction with the quantum matter which is very weak.   

It remains to study the renormalization observing that here it is not certain that the standard regularization procedures, as for example the Pauli-Villars method, will work as in the flat case. This is because of the structure of the propagators studied here which depend on mass only indirectly through the index of the $K$-functions. Thus a priority task is to find suitable methods of regularization and renormalization looking for alternative methods or adapting the well-known regularization procedures of the two-point functions \cite{Sch1,Sch2,Kel,DW,Br}.

Concluding we can say with a moderate optimism that now we have all the tools we need for calculating at least the non-gravitational effects of the massive scalar field in the presence of the gravity of the de Sitter expanding universe.

\appendix

\subsection*{Appendix A: Modified Bessel functions}
\setcounter{equation}{0} \renewcommand{\theequation}
{A.\arabic{equation}}

The modified Bessel functions $K_{\nu}(z)=K_{-\nu}(z)$ are related to the Hankel ones as 
\begin{equation}
H^{(1,2)}_{\nu}(z)=\mp\frac{2 i}{\pi}e^{\mp \frac{i}{2}\pi\nu}K_{\nu}(\mp iz)\,, \quad z\in {\Bbb R}\,,
\end{equation}
such that their Wronskian \cite{NIST} gives the identity
\begin{equation}\label{KuKu}
K_{\nu}(i s)
\stackrel{\leftrightarrow}{\partial_{s}}K_{\nu}(-is)=W[K_{\nu}(is),K_{\nu}(-is)]=\frac{i\pi}{s}\,.
\end{equation}
For $|z|\to \infty$ and any $\nu$ we have,
\begin{equation}\label{Km0}
I_{\nu}(z) \to \sqrt{\frac{\pi}{2z}}e^{z}\,, \quad K_{\nu}(z) \to K_{\frac{1}{2}}(z)=\sqrt{\frac{\pi}{2z}}e^{-z}\,.
\end{equation}

\end{document}